\documentstyle[aps,epsf,floats]{revtex}

\begin{document}

\preprint{\tighten \vbox{\hbox{FERMILAB-Conf-99/058-T} \hbox{hep-ph/9904460} }}
\baselineskip 14pt

\title{Theoretical Developments in Inclusive $B$ Decays$\,$\footnote{Invited 
talk at the American Physical Society, Division of Particles and Fields 
Conference, Jan.\ 5--9, 1999, Los Angeles, CA.}}
\author{Zoltan Ligeti}
\address{Theory Group, Fermilab, P.O.\ Box 500, Batavia, IL 60510}

\maketitle

\begin{abstract}
Some recent theoretical work on inclusive $B$ decays relevant for the model
independent determination of $|V_{ub}|$ and $|V_{cb}|$ is summarized.  The
theoretical predictions and their reliability for several differential decay
distributions in $\bar B\to X_{c,u}e\bar\nu$ and $\bar B\to X_s\gamma$ are
reviewed.  These can be used to determine certain important HQET matrix
elements.  The upsilon expansion and ways of testing it are discussed.

\end{abstract}

\section{Introduction}

In the near future, a large part of the high energy experimental program will
be devoted to testing the Cabibbo--Kobayashi--Maskawa (CKM) picture of quark
mixing and $CP$ violation by directly measuring the sides and (some) angles of
the unitarity triangle.  If the value of $\sin(2\beta)$, the $CP$ asymmetry in
$B\to J/\psi K_S$, is not too far from the CDF central value~\cite{sin2beta},
then searching for new physics at the $B$ factories will require a combination
of several precision measurements.  Particularly important are $|V_{ub}|$ and
$|V_{td}|$, which are the least precisely known elements of the CKM matrix. 
The latter will be measured hopefully in the upcoming run of the Tevatron from
the ratio of $B_s$ and $B_d$ mixing, and will not be discussed here.  This talk
is motivated by trying to understand what the chances are to

\vspace{-9pt}
\begin{itemize} \itemsep=-4pt

\item \ldots reduce (conservative) error in $|V_{cb}|$ below $\sim5\%$?
Although sometimes a smaller error is quoted already (e.g., by the Particle
Data Group), it is hard to bound model independently the possible quark-hadron
duality violation in the inclusive, and the size of $1/m_{c,b}^2$ corrections
in the exclusive determination.  

\item \ldots determine $|V_{ub}|$ with less than $\sim10\%$ error?  The
inclusive measurements require significant cuts on the available phase space;
the exclusive measurements require knowledge of the form factors.

\item \ldots reduce the theoretical uncertainties in the $B\to X_s\gamma$
photon spectrum?  The effect of the experimental cut on the photon energy needs
to be better understood, and there are subtleties in the OPE beyond leading
order.

\end{itemize}  \vspace{-9pt}

\noindent 
The theoretical reliablility of inclusive measurements can be competitive with
the exclusive ones (or even better in some cases).  For example, for the
determination of $|V_{cb}|$ model dependence enters at the same order of
$\Lambda_{\rm QCD}^2/m^2$ corrections from both the inclusive semileptonic
$\bar B\to X_c e\bar\nu$ width and the $\bar B\to D^* e\bar\nu$ rate near zero
recoil.  

Inclusive $B$ decay rates can be computed model independently in a series in
$\Lambda_{\rm QCD}/m_b$, using an operator product expansion
(OPE)~\cite{CGG,incl,MaWi}.  The $m_b\to\infty$ limit is given by $b$ quark
decay.  For most quantities of interest, this result is known including the
order $\alpha_s$ and the dominant part of the order $\alpha_s^2$ corrections. 
Observables which do not depend on the four-momentum of the hadronic final
state (e.g., total decay rate and lepton spectra) receive no correction at
order $\Lambda_{\rm QCD}/m_b$ when written in terms of $m_b$, whereas
differential rates with respect to hadronic variables (e.g., hadronic energy
and invariant mass spectra) also depend on $\bar\Lambda/m_b$, where
$\bar\Lambda$ is the $m_B-m_b$ mass difference in the $m_b\to\infty$ limit.  At
order $\Lambda_{\rm QCD}^2/m_b^2$, the corrections are parameterized by two
hadronic matrix elements, usually denoted by $\lambda_1$ and $\lambda_2$.  The
value $\lambda_2 \simeq 0.12{\rm GeV}^2$ is known from the $B^*-B$ mass
splitting.  

For inclusive $b\to q$ decay, corrections to the $m_b\to\infty$ limit are
expected to be under control in parts of phase space where several hadronic
final states are allowed to contribute with invariant masses satisfying
$m_{X_q}^2 \gtrsim m_q^2 + {\rm (few\ times)}\Lambda_{\rm QCD}
m_b$.\footnote{However, it is not true that several such states are required to
contribute; e.g., $\bar B\to X_c e\bar\nu$ decay in the small velocity limit
can be computed reliably, even though it is saturated by $D$ and $D^*$
only~\cite{SVlimit,SVdual}.}  Such observables which ``average" sufficiently
over different hadronic final states may be predicted reliably.  (We are just
beginning to learn quantitatively what sufficient averaging is.)  The major
uncertainty in these predictions is from the values of the quark masses and
$\lambda_1$, or equivalently, the values of $\bar\Lambda$ and $\lambda_1$. 
These quantities can be extracted from heavy meson decay spectra, which is the
subject of a large part of this talk.

An important theoretical subtlety is related to the fact that $\bar\Lambda$
cannot be defined unambiguously beyond perturbation theory~\cite{renormalon},
and its value extracted from data using theoretical expressions valid to
different orders in the $\alpha_s$ may vary by order $\Lambda_{\rm QCD}$. 
However, these ambiguities cancel~\cite{rencan} when one relates consistently
physical observables to one another, i.e., the formulae used to determine
$\bar\Lambda$ and $\lambda_1$ contain the same orders in the $\alpha_s$
perturbation series as those in which the so extracted values are applied to
make predictions.

\section{The HQET parameters $\bar\Lambda$ and $\lambda_1$}

The shape of the lepton energy~\cite{gremmetal,Volo,GK} or hadronic invariant
mass~\cite{FLSmass1,FLSmass2,GK} spectrum in semileptonic $\bar B\to
X_c\,\ell\,\bar\nu$ decay, and the photon spectrum in $\bar B\to
X_s\gamma$~\cite{AZ,LLMW,FLS,Bauer,KaNe} can be used to measure the heavy quark
effective theory (HQET) parameters $\bar\Lambda$ and $\lambda_1$.  Testing our
understanding of the $\bar B\to X_c e\bar\nu$ spectra is important to assess
the reliability of the inclusive determination of $|V_{cb}|$, and especially
that of $|V_{ub}|$.  Understanding the photon spectrum in $\bar B\to X_s\gamma$
is important to evaluate how precisely the total rate can be predicted in the
presence of an experimental cut on the photon energy~\cite{CLEObsg}.  Studying
these distributions is also useful to establish the limitations of models which
were built to fit the lepton energy spectrum in semileptonic decay.

The theoretical predictions are known to order $\alpha_s^2\beta_0$, where
$\beta_0=11-2n_f/3$ is the coefficient of the one-loop $\beta$-function.  This
part of the order $\alpha_s^2$ piece usually provides a reliable estimate of
the full order $\alpha_s^2$ correction, and it is straightforward to compute
using the method of Smith and Voloshin~\cite{SmVo}.  In some cases the full
$\alpha_s^2$ correction is known.  The order $\Lambda_{\rm QCD}^3/m_b^3$ terms
have also been studied~\cite{GK,FLSmass2,Bauer}, and are used to estimate the
uncertainties.  

The OPE for semileptonic (or radiative) $B$ decay does not reproduce the
physical spectrum point-by-point in regions of phase space where the hadronic
invariant mass of the final state is restricted.   For example, the maximum
electron energy for a particular hadronic final state $X$ is $E_e^{\rm
(max)}=(m_B^2-m_X^2)/2m_B$, so comparison with experimental data near the
endpoint can only be made after sufficient smearing, or after integrating over
a large enough region.  The minimal size of this region was estimated to be
around $300-500\,$MeV \cite{MaWi}.  The hadron mass spectrum cannot be
predicted point-by-point without additional assumptions, but moments of it are
calculable.  In general, higher moments of a distribution or moments over
smaller regions are less reliable than lower moments or moments over larger
regions.  The strategy is to find observables which are sensitive to
$\bar\Lambda$ and $\lambda_1$, but the deviations from $b$ quark decay are
small, so that the contributions from operators with dimension greater than 5
are not too important.

\subsection{$\bar B\to X_{\lowercase{c}} {\lowercase{e}}\bar\nu$ decay spectra}

Last year the CLEO Collaboration measured the first two moments of the hadronic
invariant mass-squared ($s_H$) distribution, $\langle s_H - \overline{m}_D^2
\rangle$ and $\langle (s_H - \overline{m}_D^2)^2 \rangle$, subject to the
constraint $E_e > 1.5\,$GeV~\cite{CLEOparams}.  Here $\overline{m}_D =
(m_D+3m_{D^*})/4$.  Each of these measurements give an allowed band on the
$\bar\Lambda - \lambda_1$ plane.  These bands are almost perpendicular, so they
give the fairly small intersection region shown in Fig.~1.  The central values
at order $\alpha_s$ are \cite{CLEOparams}
\begin{equation}\label{cleonums}
\bar\Lambda = (0.33 \pm 0.08)\, {\rm GeV}\,, \qquad 
\lambda_1 = -(0.13 \pm 0.06)\,{\rm GeV}^2\,.  
\end{equation}
The unknown order $\Lambda_{\rm QCD}^3/m_b^3$ corrections not included in this
result introduce a large uncertainty, especially for the second moment.  As a
result, the allowed range is much longer in the direction of the first moment
band than perpendicular to it; see Fig.~2.

Similar information on $\bar\Lambda$ and $\lambda_1$ can be obtained from the
lepton energy spectrum, $d\Gamma/dE_e$.  It has been measured both by demanding
only one charged lepton tag, and using a double tagged data sample where the
charge of a high momentum lepton determines whether the other lepton in the
event comes directly from semileptonic $B$ decay (primary) or from the
semileptonic decay of a $B$ decay product charmed hadron (secondary).  The
single tagged data has smaller statistical errors, but it is significantly
contaminated by secondary leptons below about $1.5\,$GeV.  Therefore,
Ref.~\cite{gremmetal} considered the observables

\noindent
\begin{minipage}[t]{8cm}
\vspace{0.1cm}
\centerline{\epsfxsize=8cm\epsffile{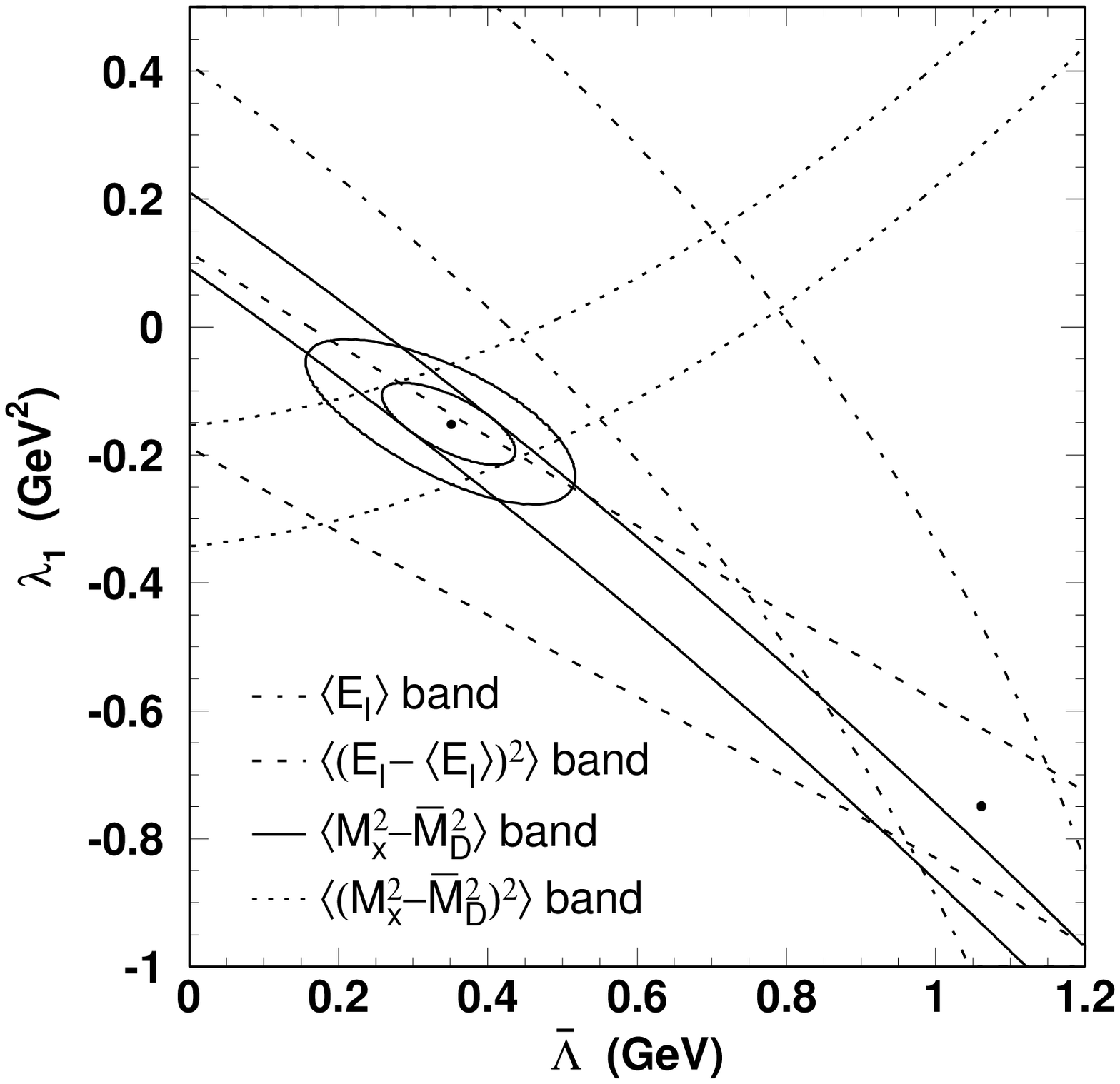}}
\small{FIG.~1.~~Bands in the $\bar\Lambda - \lambda_1$ plane defined by the 
measured first and second moments of the hadronic mass-squared and lepton 
energy distributions.  Note that $M_X^2 \equiv s_H$. 
(From Ref.~\cite{CLEOparams}.)}
\vspace{0.4cm}
\end{minipage} 
\hfill 
\begin{minipage}[t]{9cm}
\vspace{0.1cm}
\centerline{\epsfxsize=9cm\epsfysize=7cm\epsffile{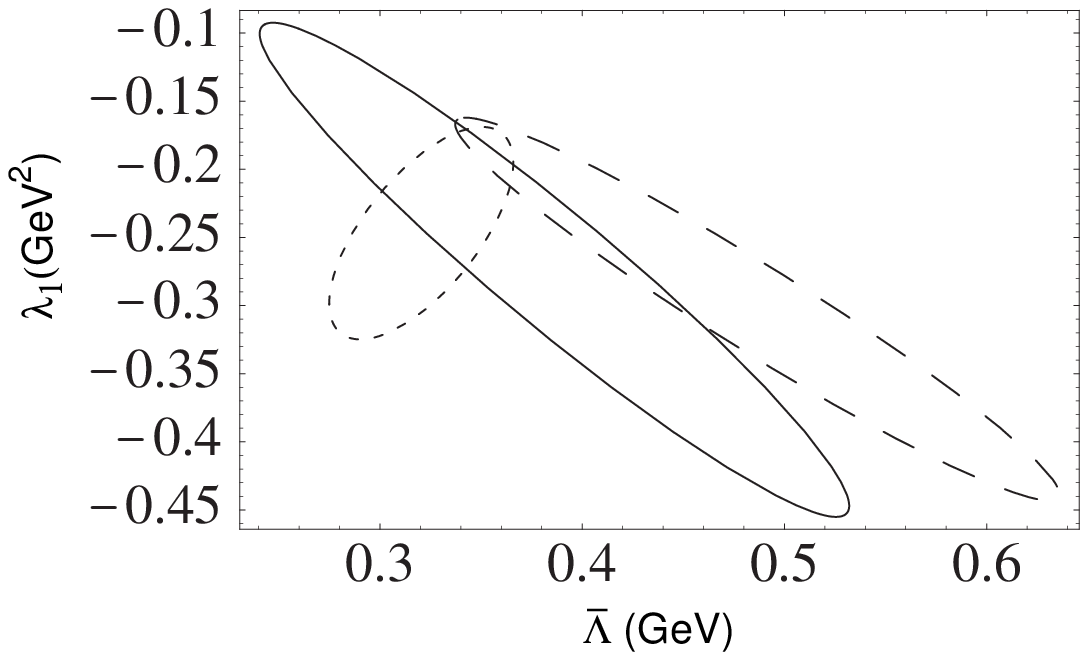}}
\small{FIG.~2.~~Estimates of the theoretical uncertainty in $\bar\Lambda$ and 
$\lambda_1$ due to unknown $1/m_b^3$ terms from the shape of the 
electron spectrum (solid ellipse), from hadronic mass moments (long dashed
ellipse), and the $\bar B\to X_s\gamma$ spectrum (short dashed ellipse; 
see Sec.~IIB).  Only the relative 
size and orientation of the ellipses are meaningful, not their
position.  (From Refs.~\cite{FLSmass2} and \cite{Bauer}.)}
\vspace{0.5cm}
\end{minipage}
\addtocounter{figure}{2}

\begin{equation}\label{R12def}
R_1 = {\displaystyle \int_{1.5\,{\rm GeV}} E_\ell\,
  {d\Gamma\over dE_\ell}\, dE_\ell \over \displaystyle
  \int_{1.5\,{\rm GeV}} {d\Gamma\over dE_\ell}\, dE_\ell }\,, \qquad
R_2 = {\displaystyle \int_{1.7\,{\rm GeV}} 
  {d\Gamma\over dE_\ell}\, dE_\ell \over \displaystyle
  \int_{1.5\,{\rm GeV}} {d\Gamma\over dE_\ell}\, dE_\ell }\,.
\end{equation}
Using the CLEO data~\cite{RoyPhD}, the central values $\bar\Lambda = 0.39\,$GeV
and $\lambda_1 = -0.19\,{\rm GeV}^2$ were obtained~\cite{gremmetal}, which
is in good agreement with Eq.~(\ref{cleonums}).

However, following Ref.~\cite{Volo}, CLEO determined the first two moments of
the spectrum, $\langle E_e \rangle$ and $\langle (E_e - \langle E_e \rangle)^2
\rangle$, without any restriction on $E_e$, using the double tagged data and an
extrapolation to $E_e < 0.6\,$GeV.  The result of this analysis is also plotted
in Fig.~1, and yields quite improbable values for $\bar\Lambda$ and
$\lambda_1$.  The extrapolation to $E_e < 0.6\,$GeV introduces unnecessary
model dependence, so this result may be less reliable than the hadronic
invariant mass analysis.  The extracted values of $\bar\Lambda$ and $\lambda_1$
are very sensitive to small systematic effects; and it seems problematic for
the exclusive models used for the extrapolation to simultaneously reproduce the
inclusive lepton spectrum and the $B$ semileptonic branching
fraction~\cite{Ryd}.

There are a number of points to emphasize regarding these results:

\vspace{-8pt}
\begin{enumerate} \itemsep=-4pt

\item Taking $\bar\Lambda$ from Eq.~(\ref{cleonums}) gives a determination of
$|V_{cb}|$ from semileptonic $B$ width with $\sim3\%$ uncertainty.  

\item The inclusive $|V_{cb}|$ seems to be slightly larger than the exclusive
(maybe not ``significantly" but ``consistently").  Can the theoretical
uncertainties be reduced by combining the exclusive and inclusive
determinations of $|V_{cb}|$?

\item Since the bands from the lepton energy spectrum and the first moment of 
the hadronic mass-squared spectrum are almost parallel, an independent 
constraint on the $\bar\Lambda - \lambda_1$ plane is needed.  
(See Section~II~B.)

\item Models do not seem to do well for $\bar B\to D^{**}e\bar\nu$ and $\bar
B\to D^{(*)}\pi e\bar\nu$ --- relying on them may be dangerous.  Composition of
semileptonic $B$ decay seems to be not really understood yet.  (See next
subsection.)

\end{enumerate} \vspace{-8pt}

\subsubsection{Semileptonic $B$ decays to excited charmed mesons}

In the heavy quark symmetry limit~\cite{HQS}, the spin and parity of the light
degrees of freedom in a heavy meson are conserved.  In the charm sector, the
ground state is the $(D,\, D^*)$ doublet of heavy quark spin symmetry with
spin-parity of the light degrees of freedom $s_\ell^{\pi_\ell} = \frac12^-$. 
The four lightest excited states, sometimes referred to as $D^{**}$, are the
$(D_0^*,\, D_1^*)$ doublet with $s_\ell^{\pi_\ell} = \frac12^+$ and the
$(D_1,\, D_2^*)$ doublet with $s_\ell^{\pi_\ell} = \frac32^-$.  The $D_1$ and
$D_2^*$ have been observed with masses near $2.42$ and $2.46\,$GeV,
respectively, and width around $20\,$MeV.  States in the
$s_l^{\pi_l}=\frac12^+$ doublet can decay into $D^{(*)}\pi$ in an $s$-wave, and
so they are expected to be much broader than the $D_1$ and $D_2^*$ which can
only decay in a $d$-wave.  (An $s$-wave decay for the $D_1$ is forbidden by
heavy quark spin symmetry \cite{IWprl}.)  The first observation of the $D_1^*$
state with mass and width about $2.46\,$GeV and $290\,$MeV, respectively, was
reported at this Conference~\cite{CLEObroad}.

$\bar B\to D_1 e\bar\nu$ and $\bar B\to D_2^* e\bar\nu$ account for sizable
fractions of semileptonic $B$ decays, and are probably the only three-body
semileptonic $B$ decays (other than $\bar B\to D^{(*)} e\bar\nu$) whose
differential decay distributions will be precisely measured.  ALEPH and CLEO
measured recently, with some assumptions, ${\cal B}(\bar B\to D_1 e\bar\nu_e) =
(6.0\pm1.1) \times 10^{-3}$~\cite{AlephCleo}, while $\bar B\to D_2^*
e\bar\nu_e$ has not been seen yet.  Heavy quark symmetry implies that in the
$m_Q\to\infty$ limit ($Q=c,b$), matrix elements of the weak currents between a
$B$ meson and an excited charmed meson vanish at zero recoil.  However, in some
cases at order $\Lambda_{\rm QCD}/m_Q$ these matrix elements are not zero. 
Since most of the phase space for semileptonic $B$ decay to excited charmed
mesons is near zero recoil, $1 < v \cdot v' \lesssim 1.3$, $\Lambda_{\rm
QCD}/m_Q$ corrections can be very important.  

The matrix elements of the weak currents between $B$ mesons and $D_1$ or
$D_2^*$ mesons are conventionally parameterized in terms of a set of eight form
factors $f_i$ and $k_i$~\cite{LLSW}.  At zero recoil only $f_{V_1}$, defined by
\begin{equation}\label{formf1}
\langle D_1(v',\epsilon)|\, \bar c\,\gamma^\mu\,b\, |B(v)\rangle =
  \sqrt{m_{D_1}\,m_B}\, [ f_{V_1}\, \epsilon^{*\mu} 
  + (f_{V_2} v^\mu + f_{V_3} v'^\mu)\, (\epsilon^*\cdot v) ] \,,
\end{equation}
can contribute to the matrix elements.  In the $m_Q\to\infty$ limit, $f_i$ and
$k_i$ are given in terms of a single Isgur-Wise function,
$\tau(w)$~\cite{IWsr}.  Heavy quark symmetry does not fix $\tau(1)$, since
$f_{V_1} = (1-w^2)\,\tau(w) + {\cal O}(1/m_Q)$, and so $f_{V_1}(1) = 0$ in the
infinite mass limit independent of the value of $\tau(1)$.  

At order $1/m_Q$ several new Isgur-Wise functions occur, together with the
parameter $\bar\Lambda'$, which is the analog of $\bar\Lambda$ for the $(D_1,\,
D_2^*)$ doublet.  $\bar\Lambda' - \bar\Lambda \simeq 0.39\,$GeV follows from
the measured meson masses~\cite{LLSW}.  At order $1/m_Q$, $f_{V_1}(1)$ is no
longer zero, but it can be written in terms of $\bar\Lambda'-\bar\Lambda$ and
the Isgur-Wise function $\tau(w)$ evaluated at zero recoil~\cite{LLSW},
\begin{equation}\label{fV1}
\sqrt{6}\, f_{V_{1}}(1) = 
  - {4 (\bar\Lambda'-\bar\Lambda) \over m_c}\, \tau (1)\,.
\end{equation}
This relation means that at zero recoil heavy quark symmetry gives some model
independent information about the $1/m_Q$ corrections, similar to Luke's
theorem~\cite{Luke} for the decay into the ground state $(D,\, D^*)$ doublet. 

Since the allowed kinematic range for $\bar B\to D_1 e\bar\nu_e$  and $\bar
B\to D_2^* e\bar\nu_e$ decay are fairly small ($ 1 < w \lesssim 1.3$), and
there are some constraints on the $1/m_Q$ corrections at zero recoil, it is
useful to consider the decay rates expanded in powers of $w-1$.  The general
structure of the expansion of $d\Gamma/dw$ is elucidated
schematically below,
\begin{eqnarray}\label{schematic}
{d\Gamma_{D_1}^{(\lambda=0)}\over dw} &\sim&
  \sqrt{w^2-1}\, \Big[ ( 0
  + 0\,\varepsilon + \varepsilon^2 + \ldots )
  + (w-1)\, ( 0 + \varepsilon + \ldots ) 
  + (w-1)^2\, ( 1 + \varepsilon + \ldots ) + \ldots \Big] \,, 
  \nonumber\\*
{d\Gamma_{D_1}^{(|\lambda|=1)}\over dw} &\sim&
  \sqrt{w^2-1}\, \Big[ ( 0 
  + 0\,\varepsilon + \varepsilon^2 + \ldots ) 
  + (w-1)\, ( 1 + \varepsilon + \ldots )
  + (w-1)^2\, ( 1 + \varepsilon + \ldots ) + \ldots \Big] \,, \nonumber\\*
{d\Gamma_{D_2^*}^{(|\lambda|=0,1)}\over dw} &\sim&
  (w^2-1)^{3/2} \Big[ ( 1 + \varepsilon + \ldots ) 
  + (w-1) ( 1 + \varepsilon + \ldots ) + \ldots \Big] \,. 
\end{eqnarray}
Here $\lambda$ is the helicity of the $D_1$ or $D_2^*$, and $\varepsilon^n$
denotes a term of order $(\Lambda_{\rm QCD}/m_Q)^n$.  The zeros in
Eq.~(\ref{schematic}) are consequences of heavy quark symmetry. 

\begin{table}[t] 
\begin{tabular}{c|ccc}  
~~~Approximation~~~  &  $R=\Gamma_{D_2^*}\big/\Gamma_{D_1}$  &  
  $\tau(1)\, \bigg[\displaystyle {6.0\times10^{-3} \over 
    {\cal B}(\bar B\to D_1\,e\,\bar\nu_e)} \bigg]^{1/2}$  &
  $\Gamma_{D_1+D_2^*+D_1^*+D_0^*}\big/\Gamma_{D_1}$~~~  \\[8pt] \hline 
B$_\infty$  &  $1.65$  &  $1.24$  &  $4.26$  \\
B$_1$  &  $0.52$  &  $0.71$  &  $2.55$  \\
B$_2$  &  $0.67$  &  $0.75$  &  $2.71$  \\
\end{tabular} \vspace{4pt}
\caption[6]{Predictions for $\Gamma_{D_2^*}/\Gamma_{D_1}$, $\tau(1)$, and 
$\Gamma_{D_1+D_2^*+D_1^*+D_0^*}/\Gamma_{D_1}$.  (From Ref.~\cite{LLSW}.)}
\end{table}

Table~I summarizes some of the more important predictions.  The first row
($B_\infty$) shows the infinite mass limit.  Approximations $B_1$ and $B_2$ are
two different ways of treating unknown $1/m_Q$ corrections, and may give an
indication of the uncertainty at this order.   One of the most interesting
predictions is that the $B\to D_1$ decay rate should be larger than $B\to
D_2^*$, contrary to the infinite mass limit, but in agreement with the data.  A
number of other predictions for different helicity amplitudes, factorization of
$\bar B\to (D_1,\, D_2^*)\pi e\bar\nu$ decay, sum rules, etc., are presented in
Ref.~\cite{LLSW} (see also \cite{MNfact}).  The main points are: 

\vspace{-8pt}
\begin{enumerate} \itemsep=-4pt

\item At zero recoil, order $1/m_Q$ contributions to semileptonic $B\to
D_1,\, D_2^*$ decays (any excited charmed meson with $+$ parity) are determined
by the $m_Q\to\infty$ Isgur-Wise function and known hadron mass splittings.

\item Decay spectra can be predicted near zero recoil, including the $1/m_Q$
corrections, with reasonable assumptions.

\item Test heavy quark symmetry for $B$ decays to excited charmed mesons, where
the $1/m_Q$ terms are sometimes the leading contributions.  Large $1/m_Q$
corrections to some predictions can be checked against data.

\item Better understanding of inclusive$\,=\sum\,$exclusive in semileptonic $B$
decay.  If semileptonic $B$ decays into the four $D^{**}$ states indeed account
for less than 2\% of the $B$ width, then about another 2\% of $B$ decays must
be semileptonic decays to higher mass excitations or nonresonant channels.

\item Constrain the validity of models constructed to fit the decays to the
ground state final state (most predict, for example, larger $B\to D_2^*$ than
$B\to D_1$ semileptonic rate).

\end{enumerate} \vspace{-8pt}

\subsection{$\bar B\to X_{\lowercase{s}} \gamma$ photon spectrum}

Comparison of the measured weak radiative $\bar B\to X_s\gamma$ decay rate with
theory is an important test of the standard model.  In contrast to the decay
rate itself, the shape of the photon spectrum is not expected to be sensitive
to new physics, but it can nevertheless provide important information.  First
of all, studying the photon spectrum is important for understanding how
precisely the total rate can be predicted in the presence of an experimental
cut on the photon energy~\cite{CLEObsg}, which is needed for a model
independent interpretation of the resulting decay rate.  Secondly, moments of
the photon spectrum may be used to measure the HQET parameters $\bar\Lambda$
and $\lambda_1$~\cite{AZ,LLMW,FLS,Bauer,KaNe}.

An important observable is
\begin{equation}\label{moment1}
\overline{(1 - x_B)} \Big|_{x_B > 1 - \delta} = 
  {\displaystyle \int_{1-\delta}^1 dx_B\, (1-x_B)\, \frac{d\Gamma}{dx_B} \over
   \displaystyle \int_{1-\delta}^1 dx_B\, \frac{d\Gamma}{dx_B} } \,,
\end{equation}
where $x_B = 2E_\gamma / m_B$ is the rescaled photon energy.  The parameter 
$\delta = 1 - 2E_\gamma^{\rm min}/m_B$ corresponds to the experimental lower 
cut on $E_\gamma$, and it has to satisfy $\delta > \Lambda_{\rm QCD}/m_B$; 
otherwise nonperturbative effects are not under control.  
It is straightforward to show that~\cite{LLMW}
\begin{equation}\label{beauty}
\overline{(1 - x_B)} \Big|_{x_B > 1-\delta} =
  {\bar\Lambda\over m_B} + \bigg(1-{\bar\Lambda\over m_B}\bigg)\,
  \int_{1-\delta}^1 dx_b\, 
  (1-x_b)\, \frac1{\Gamma_0}\, \frac{d\Gamma}{dx_b}
 - {\bar\Lambda\over m_B}\, \delta(1-\delta)\, \frac1{\Gamma_0}\,
  \frac{d\Gamma}{dx_b} \bigg|_{x_b = 1-\delta} + \ldots \,,
\end{equation}
where $x_b = 2E_\gamma / m_b$, and $d\Gamma/dx_b$ is the photon spectrum in
$b$ quark decay.  It has recently been computed away from the endpoint ($x_b =
1$) to order $\alpha_s^2\beta_0$~\cite{LLMW}.  $\Gamma_0 =
G_F^2\,|V_{tb}V_{ts}^*|^2\,\alpha_{\rm em}\,C_7^2\, m_b^5 / 32\pi^4$ is the
contribution of the tree level matrix element of $O_7 (\sim \bar s_L
\sigma_{\mu\nu} F^{\mu\nu} b_R)$ to the $\bar B\to X_s\gamma$ decay rate.  All 
terms but the first one on the right-hand-side of Eq.~(\ref{beauty}) have
perturbative expansions which begin at order $\alpha_s$.  The ellipses denote
contributions of order $(\Lambda_{\rm QCD}/m_B)^3$, $\alpha_s(\Lambda_{\rm
QCD}/m_B)^2$, and $\alpha_s^2$ terms not enhanced by $\beta_0$, but it does not
contain contributions of order $(\Lambda_{\rm QCD}/m_B)^2$ or additional terms
of order $\alpha_s(\Lambda_{\rm QCD}/m_B)$.  Terms in the operator product
expansion proportional to $\lambda_{1,2}/m_b^2$ enter precisely in the form so
that they are absorbed in $m_B$ in Eq.~(\ref{beauty})~\cite{AZ}.  

A determination of $\bar\Lambda$ is straightforward using Eq.~(\ref{beauty}). 
The left hand side is directly measurable, while the quantities entering on the
right-hand-side are presented in Ref.~\cite{LLMW}.  The CLEO data in the region
$E_\gamma > 2.1\,$GeV~\cite{CLEObsg} yields the central values
$\bar\Lambda_{\alpha_s^2\beta_0} \simeq 270\,$MeV and $\bar\Lambda_{\alpha_s}
\simeq 390\,$MeV.  Uncertainties due to the unknown order $\Lambda_{\rm
QCD}^3/m_b^3$ terms in the OPE give the short dashed ellipse in
Fig.~2~\cite{Bauer}, whose major axis is roughly perpendicular to those from
$\bar B\to X_c e\bar\nu$.  This is why it is important to determine
$\bar\Lambda$ and $\lambda_1$ from both analyses.  The potentially most serious
uncertainty is from both nonperturbative and perturbative terms that are
singular as $x_b\to 1$ and sum into a shape function~\cite{shapefn} that
modifies the spectrum near the endpoint.  For sufficiently large $\delta$ these
effects are not important.  They have been estimated in Refs.~\cite{Bauer,KaNe}
using phenomenological models.  Whether these effects are small in a certain
range of $\delta$ can be tested experimentally by checking if the extracted
value of $\bar\Lambda$ is independent of $\delta$.  This would also improve our
confidence that the total decay rate in the region $x_B>1-\delta$ can be
predicted model independently.

There have been other important developments for the $\bar B\to X_s\gamma$
decay rate.  The next-to-leading order computation of the total rate has been
completed~\cite{Misiak,match,fourquark,matrixel}, reducing the theoretical
uncertainties significantly.  It was realized by Voloshin that beyond leading
order there are terms in the OPE for the decay rate which are suppressed only
by $\Lambda_{\rm QCD}^2/m_c^2$ instead of $\Lambda_{\rm
QCD}^2/m_b^2$~\cite{1/mc2}.  In fact, there is a series of contributions of the
form $(\Lambda_{\rm QCD}^2/m_c^2)\, (\Lambda_{\rm QCD} m_b/m_c^2)^n$, which
gives for $n=0$ a calculable correction of $\delta\Gamma/\Gamma = - (C_2/9C_7)
(\lambda_2/m_c^2) \simeq 2.5\%$.  It is also known that there are uncalculable
contributions suppressed by $\alpha_s$ but not by $\Lambda_{\rm QCD}/m_b$ from
photon coupling to light quarks, for which there is no OPE~\cite{KLP}.  While
no correction larger than a few percent has been identified, a better
understanding of the nonperturbative contributions would be desirable once the
four-quark operators are included.

\section{$|V_{\lowercase{ub}}|$ from the $\bar B\to X_{\lowercase{u}}
{\lowercase{e}}\bar\nu$ hadron mass spectrum}

The traditional method for extracting $|V_{ub}|$ involves a study of the
electron energy spectrum in inclusive semileptonic $B$ decay.  Electrons with
energies in the endpoint region $m_B/2 > E_e > (m_B^2-m_D^2)/2m_B$ (in the $B$
rest frame, and neglecting the pion mass) must arise from $b \to u$ transition.
Since the size of this region is only about $300\,$MeV, at the present time it
is not known how to make a model independent prediction for the spectrum in
this region.  Another possibility for extracting $|V_{ub}|$ is based on
reconstructing the neutrino momentum.  The idea is to reconstruct $p_{\bar\nu}$
and infer the invariant mass-squared of the hadronic final state, $s_H = (p_B -
q)^2$, where $q = p_e + p_{\bar\nu}$.  Semileptonic $B$ decays satisfying $s_H
< m_D^2$ must come from $b \to u$ transition~\cite{FLW,DiUr,oldmass}.  The
first analyses of LEP data utilizing this idea have been performed
recently~\cite{Vubmass}.

Both the invariant mass region, $s_H < m_D^2$, and the electron endpoint
region, $m_B/2 > E_e > (m_B^2-m_D^2)/2m_B$, receive contributions from hadronic
final states with invariant masses between $m_\pi$ and $m_D$.  However, for the
electron endpoint region the contribution of states with masses nearer to $m_D$
is strongly suppressed kinematically.  In fact, in the ISGW model~\cite{ISGW}
the electron endpoint region is dominated by the $\pi$ and the $\rho$ with
higher mass states making a small contribution, and this region includes only
of order 10\% of the $\bar B\to X_u e\bar\nu$ rate.  The situation is very
different for the low invariant mass region, $s_H < m_D^2$.  Now all states
with invariant masses up to $m_D$ contribute without any preferential weighting
towards the lowest mass ones.  In this case the ISGW model suggests the $\pi$
and the $\rho$ comprise only about a quarter of the $B$ semileptonic decays to
states with $s_H < m_D$, and only of order 10\% of the $\bar B\to X_u e\bar\nu$
rate is excluded from this region.  Consequently, it is much more likely that
the first few terms in the OPE provide an accurate description of $B$
semileptonic decay in the region $s_H < m_D^2$ than in the region $E_e >
(m_B^2-m_D^2)/2m_B$.  Combining the invariant mass constraint with a modest cut
on the electron energy will not destroy this conclusion.

Let us first consider the contribution of dimension three operators in the OPE
to the hadronic mass-squared spectrum.  This is equivalent to $b$ quark decay,
and implies a result for $d\Gamma / dE_0\, ds_0$ (where $E_0 = p_b \cdot
(p_b-q)/m_b$ and $s_0 = (p_b-q)^2$ are the energy and invariant mass of the
strongly interacting partons arising from the $b$ quark decay) that is
straightforward to calculate to order $\alpha_s^2\beta_0$~\cite{FLW}.  Even at
this leading order in the OPE there are important nonperturbative effects that
come from the difference between $m_b$ and $m_B$.  The most significant effect
comes from $\bar\Lambda$, and including only it ({\it i.e.}, neglecting
$\lambda_{1,2}$), the hadronic invariant mass $s_H$ is related to $s_0$ and
$E_0$ by~\cite{FLSmass1}
\begin{equation}
s_H = s_0 + 2 \bar\Lambda E_0 + \bar\Lambda^2 + \ldots \,.
\end{equation}
Changing variables from $(s_0, E_0)$ to $(s_H, E_0)$ and integrating $E_0$
over the range
\begin{equation}
\sqrt{s_H} - \bar\Lambda < E_0 <
  {1\over 2m_B}\, (s_H - 2\bar\Lambda m_B + m_B^2),
\end{equation}
gives $d\Gamma/ds_H$, where $\bar\Lambda^2<s_H<m_B^2$.  Feynman diagrams with 
only a $u$ quark in the final state contribute at $s_0 = 0$, which corresponds 
to the region $\bar\Lambda^2 < s_H < \bar\Lambda m_B$.

Although $d\Gamma/ds_H$ is integrable in perturbation theory, it has a double
logarithmic singularity at $s_H = \bar\Lambda m_B$.  At higher orders in
perturbation theory, increasing powers of $\alpha_s \ln^2[(s_H - \bar\Lambda
m_B) / m_B^2]$ appear in the invariant mass spectrum.  Therefore,
$d\Gamma/ds_H$ in the vicinity of $s_H = \bar\Lambda m_B$ is hard to predict
reliably even in perturbation theory.  (In the region $s_H \lesssim \bar\Lambda
m_B$ nonperturbative effects are also important.)  The behavior of the spectrum
near $s_H = \bar\Lambda m_B$ becomes less important for observables that
average over larger regions of the spectrum, such as $d\Gamma/ds_H$ integrated
over $s_H<\Delta^2$, with $\Delta^2$ significantly greater than $\bar\Lambda
m_B$.  Fig.~3 shows the fraction of $\bar B\to X_u e\bar\nu$ events in the
region $s_H < \Delta^2$ as a function of $\Delta^2$ for three different values
of $\bar\Lambda$.  For a certain value of $\bar\Lambda$, Fig.~3 together with
Eq.~(\ref{Vub}) can be used to extract $|V_{ub}|$ from data, up to the
nonperturbative effects discussed next.

\begin{figure}[t]
\centerline{\epsfysize=7cm\epsffile{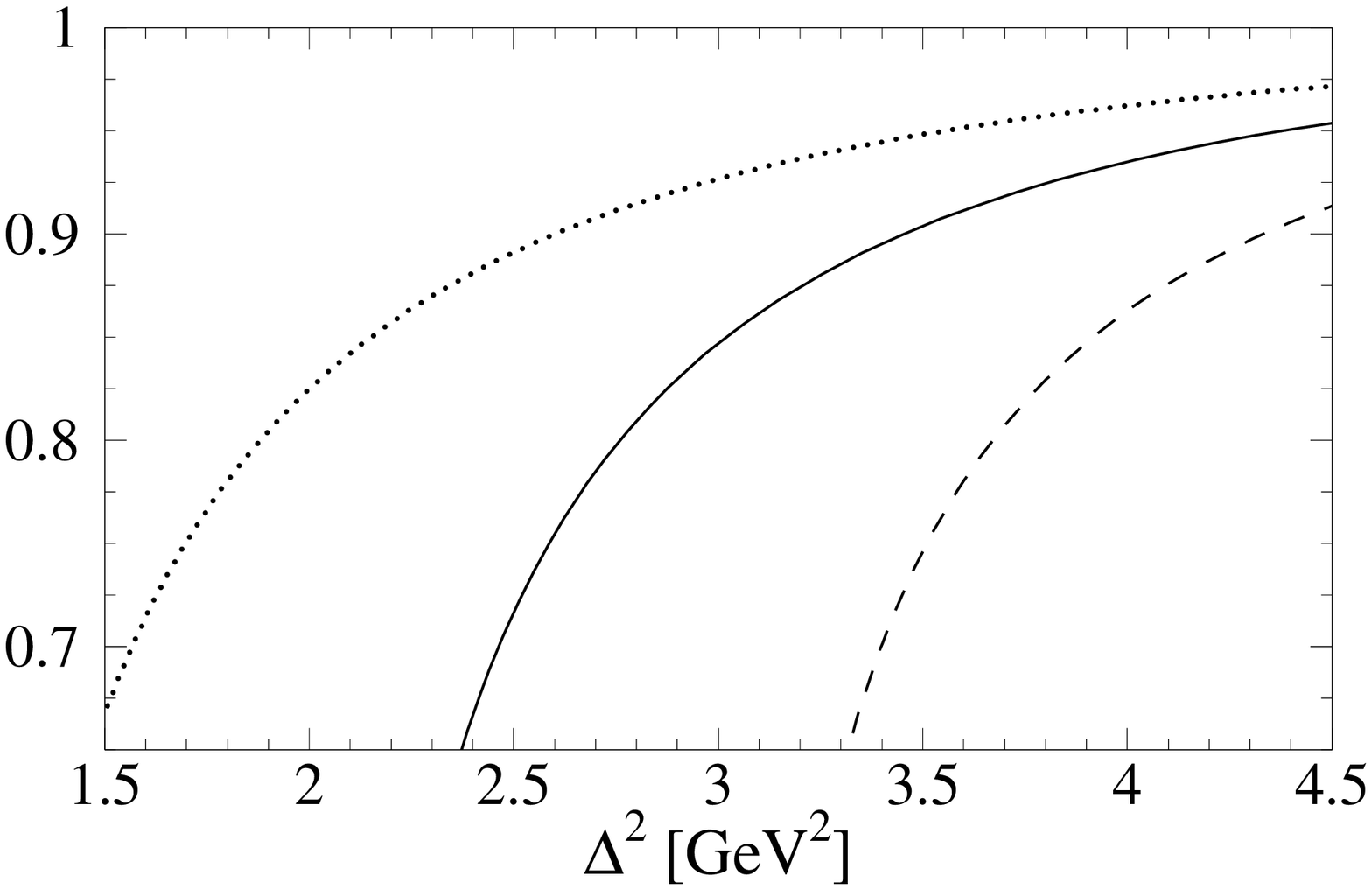}}
\caption[]{Fraction of $\bar B\to X_u e\bar\nu$ decays with $s_H < \Delta^2$, 
for $\bar\Lambda = 0.2\,$GeV (dotted curve), $0.4\,$GeV (solid curve), and 
$0.6\,$GeV (dashed curve).  Note that $m_D^2 = 3.5\,{\rm GeV}^2$.
Nonperturbative effects not in $\bar\Lambda$ are neglected.
(From Ref.~\cite{FLW}.) }
\end{figure}

In the low mass region, $s_H \lesssim \bar\Lambda m_B$, nonperturbative
corrections from higher dimension operators in the OPE are very important. 
Just as in the case of the electron endpoint region in semileptonic $B$ decay
or the photon energy endpoint region in radiative $B$ decay, the most singular
terms can be identified and summed into a shape function.  These shape
functions depend on the same infinite set of matrix elements.  Since
$\bar\Lambda m_b\approx2\,{\rm GeV}^2$ is not too far from $m_D^2$, it is
necessary to estimate the influence of the nonperturbative effects on the
fraction of $B$ decays with $s_H < m_D^2$.  It is difficult to estimate this
model independently, but upper bounds can be derived on the fraction of $\bar
B\to X_u e\bar\nu$ events with $s_H > \Delta^2$ assuming that the shape
function is positive~\cite{FLW}.  In the ACCMM model \cite{ACCMM} with
reasonable parameters, the shape function causes a small (i.e., $\sim4\%$ with
$\bar\Lambda=0.4\,$GeV, and perturbative QCD corrections neglected) fraction of
the events to have $s_H>m_D^2$ \cite{FLW}.  This suggests that sensitivity to
unknown higher dimension operators in the OPE will probably not give rise to a
large uncertainty in $|V_{ub}|$ if it is determined from the hadronic invariant
mass spectrum in the region $s_H<m_D^2$.  If experimental resolution forces one
to consider a significantly smaller region, then the sensitivity to higher
dimension operators increases rapidly.

In summary, to extract $|V_{ub}|$ from $d\Gamma/ds_H$ with small theoretical
uncertainty, one needs to:

\vspace{-9pt}
\begin{enumerate} \itemsep=-4pt

\item move the experimental cut $\Delta$ as close to $m_D$ as possible;

\item determine $\bar\Lambda$ (at order $\alpha_s^2$) with $\lesssim 50\,$MeV
uncertainty.

\end{enumerate} \vspace{-9pt}
\noindent
Then a determination of $|V_{ub}|$ with $\sim 10\%$ theoretical uncertainty 
seems feasible.

\section{Upsilon expansion}

The main uncertainties in the theoretical predictions for inclusive $B$ decay
rates, e.g., $\bar B\to X_u e\bar\nu$, arise from the $m_b^5$ dependence on the
$b$ quark mass and the bad behavior of the series of perturbative corrections
when it is written in terms of the pole mass.  In fact, only the product of
these quantities is unambiguous, but perturbative multi-loop calculations are
most comfortably done in terms of the pole mass.  Of course, one would like to
eliminate the quark mass altogether from the predictions in favor of a physical
observable.  Here we present a new method of eliminating $m_b$ in terms of the
$\Upsilon(1S)$ meson mass~\cite{upsexp} (instead of $m_B$ and $\bar\Lambda$
discussed in Sec.~II).

Let us consider the inclusive $\bar B\to X_u e\bar\nu$ decay rate~\cite{LSW}.  
At the scale $\mu = m_b$,
\begin{equation}\label{bupole}
\Gamma(B\to X_u e\bar\nu) = {G_F^2 |V_{ub}|^2\over 192\pi^3}\, m_b^5\,
  \bigg[ 1 - 2.41 {\alpha_s\over\pi}\, \epsilon 
  - 3.22 {\alpha_s^2\over\pi^2} \beta_0\, \epsilon^2 
  - 5.18 {\alpha_s^3\over\pi^3} \beta_0^2\, \epsilon^3 - \ldots 
  - {9\lambda_2 - \lambda_1 \over 2m_b^2} + \ldots \bigg] . 
\end{equation}
The variable $\epsilon \equiv 1$ denotes the order in the modified expansion. 
The complete order $\alpha_s^2$ calculation was done recently~\cite{TvR}, and
the result is about 90\% of the $\alpha_s^2\beta_0$ part.  In comparison, the
expansion of the $\Upsilon(1S)$ mass in terms of $m_b$~\cite{Upsmass} has a
different structure,
\begin{equation}\label{upsmass}
{m_\Upsilon \over 2m_b} = 1 - {(\alpha_s C_F)^2\over8} \bigg\{ 1 \epsilon 
+ {\alpha_s\over\pi} \bigg[\bigg( \ell + \frac{11}6\bigg) \beta_0 
  - 4 \bigg] \epsilon^2 
+ \bigg({\alpha_s\beta_0\over2\pi}\bigg)^2 
  \bigg( 3\ell^2 +9\ell +2\zeta(3)+\frac{\pi^2}6+\frac{77}{12}\bigg) 
  \epsilon^3 + \ldots \bigg\}\, ,
\end{equation}
where $\ell=\ln[\mu/(m_b\alpha_s C_F)]$ and $C_F=4/3$.  In this expansion we
assigned to each term one less power of $\epsilon$ than the power of
$\alpha_s$, because as we will sketch below, this is the consistent way of
combining Eqs.~(\ref{bupole}) and (\ref{upsmass}).  It is also convenient to
choose the same renormalization scale $\mu$.  The prescription of counting
$[\alpha_s(m_b)]^n$ in $B$ decay rates as order $\epsilon^n$, and
$[\alpha_s(m_b)]^n$ in $m_\Upsilon$ as order $\epsilon^{n-1}$ is called the
upsilon expansion.  Note that it combines different orders in the $\alpha_s$
perturbation series in Eqs.~(\ref{bupole}) and (\ref{upsmass}).

The theoretical consistency of the upsilon expansion was shown at large orders
for the terms containing the highest possible power of $\beta_0$, and to order
$\epsilon^2$ including non-Abelian contributions.  An explicit calculation
using the Borel transform of the static quark potential~\cite{ugo} shows that
the coefficient of the order $\alpha_s^{n+2}$ term in Eq.~(\ref{upsmass}) of
the form $(\ell^n+ \ell^{n-1}+ \ldots+1)$ exponentiates to give $\exp(\ell) =
\mu/(m_b\alpha_s C_F)$, and corrects the mismatch of the power of $\alpha_s$
between the two series.  This is also needed for the cancellation of the
renormalon ambiguities in the energy levels as given by $2m_b$ plus the
potential and kinetic energies~\cite{andre,Beneke}.  The infrared sensitivity
of Feynman diagrams can be studied by introducing a fictitious infrared cutoff
$\lambda$.  The infrared sensitive terms are nonanalytic in $\lambda^2$, such
as $(\lambda^2)^{n/2}$ or $\lambda^{2n}\ln\lambda^2$, and arise from the
low-momentum part of Feynman diagrams.  Diagrams which are more infrared
sensitive, i.e., have contributions $(\lambda^2)^{n/2}$ or
$\lambda^{2n}\ln\lambda^2$ for small values of $n$, are expected to have larger
nonperturbative contributions.  Linear infrared sensitivity, i.e., terms of
order $\sqrt{\lambda^2}$, are a signal of $\Lambda_{\rm QCD}$ effects,
quadratic sensitivity, i.e., terms of order $\lambda^2\ln \lambda^2$ are a
signal of $\Lambda_{\rm QCD}^2$ effects, etc.  From Refs.~\cite{SAZ} and
\cite{Beneke} follows that the linear infrared sensitivity cancels in the
upsilon expansion to order $\epsilon^2$ (probably to all orders as well, but
the demonstration of this appears highly non-trivial).

Substituting Eq.~(\ref{upsmass}) into Eq.~(\ref{bupole}) and collecting terms
of a given order in $\epsilon$ gives~\cite{upsexp}
\begin{equation}\label{buups}
\Gamma(\bar B\to X_u e\bar\nu) = {G_F^2 |V_{ub}|^2\over 192\pi^3}\,
  \bigg({m_\Upsilon\over2}\bigg)^5\, \bigg[ 1 - 0.115\epsilon
  - 0.035_{\rm BLM} \epsilon^2 - 0.005_{\rm BLM} \epsilon^3 
  - {9\lambda_2 - \lambda_1 \over 2(m_\Upsilon/2)^2} + \ldots \bigg] ,
\end{equation}
where the BLM~\cite{BLM} subscript indicates that only the corrections
proportional to the highest power of $\beta_0$ have been kept.  The complete
order $\epsilon^2$ term is $-0.041 \epsilon^2$~\cite{TvR}.  The perturbation
series, $1 - 0.115\epsilon - 0.035_{\rm BLM}\epsilon^2 - 0.005_{\rm
BLM}\epsilon^3$, is far better behaved than the series in Eq.~(\ref{bupole}),
$1 - 0.17\epsilon - 0.13_{\rm BLM}\epsilon^2 - 0.12_{\rm BLM}\epsilon^3$, or
the series expressed in terms of the $\overline{\rm MS}$ mass,
$1+0.30\epsilon+0.19_{\rm BLM}\epsilon^2+0.05_{\rm BLM}\epsilon^3$.  The
uncertainty in the decay rate using Eq.~(\ref{buups}) is much smaller than that
in Eq.~(\ref{bupole}), both because the perturbation series is better behaved,
and because $m_\Upsilon$ is better known (and better defined) than $m_b$.  The
relation between $|V_{ub}|$ and the total semileptonic $\bar B\to X_u e\bar\nu$
decay rate is~\cite{upsexp}
\begin{equation}\label{Vub}
|V_{ub}| = (3.06 \pm 0.08 \pm 0.08) \times 10^{-3} 
  \left( {{\cal B}(\bar B\to X_u e\bar\nu)\over 0.001}
  {1.6\,{\rm ps}\over\tau_B} \right)^{1/2} ,
\end{equation}
The first error is obtained by assigning an uncertainty in Eq.~(\ref{buups})
equal to the value of the $\epsilon^2$ term and the second is from assuming a
$100\,$MeV uncertainty in Eq.~(\ref{upsmass}).  The scale dependence of
$|V_{ub}|$ due to varying $\mu$ in the range $m_b/2< \mu <2m_b$ is less than
1\%.  The uncertainty in $\lambda_1$ makes a negligible contribution to the
total error.  Of course, it is unlikely that ${\cal B}(\bar B\to X_u e\bar\nu)$
will be measured without significant experimental cuts, for example, on the
hadronic invariant mass (see Sec.~III), but this method should reduce the
uncertainties in such analyses as well.

The $\bar B\to X_c e\bar\nu$ decay depends on both $m_b$ and $m_c$.  It is
convenient to express the decay rate in terms of $m_\Upsilon$ and $\lambda_1$
instead of $m_b$ and $m_c$, using Eq.~(\ref{upsmass}) and
\begin{equation}\label{mbmc}
m_b - m_c = \overline{m}_B - \overline{m}_D + \bigg( 
  {\lambda_1\over 2\overline{m}_B} - {\lambda_1\over 2\overline{m}_D} \bigg) 
  + \ldots \,,
\end{equation}
where $\overline{m}_B = (3m_{B^*}+m_B)/4=5.313\,$GeV and $\overline{m}_D =
(3m_{D^*}+m_D)/4=1.973\,$GeV.  We then find
\begin{equation}\label{bcups}
\Gamma(\bar B\to X_c e\bar\nu) = {G_F^2 |V_{cb}|^2\over 192\pi^3}
  \bigg({m_\Upsilon\over2}\bigg)^5\, 0.533 
  \times \big[ 1 - 0.096\epsilon - 0.029_{\rm BLM}\epsilon^2
  - (0.28\lambda_2 + 0.12\lambda_1)/{\rm GeV}^2 \big] \,,
\end{equation}
where the phase space factor has also been expanded in $\epsilon$.  For
comparison, the perturbation series in this relation when written in terms of
the pole mass is $ 1- 0.12\epsilon- 0.06\epsilon^2 -\ldots$~\cite{LSW}. 
Equation~(\ref{bcups}) implies~\cite{upsexp}
\begin{equation}\label{Vcb}
|V_{cb}| = (41.6 \pm 0.8 \pm 0.7 \pm 0.5) \times 10^{-3}
  \times \eta_{\rm QED} \left( {{\cal B}(\bar B\to X_c e\bar\nu)\over0.105}\,
  {1.6\,{\rm ps}\over\tau_B}\right)^{1/2} ,
\end{equation}
where $\eta_{\rm QED}\sim1.007$ is the electromagnetic radiative correction. 
The uncertainties come from assuming an error in Eq.~(\ref{bcups}) equal to the
$\epsilon^2$ term, a $0.25\,{\rm GeV}^2$ error in $\lambda_1$, and a $100\,$MeV
error in Eq.~(\ref{upsmass}), respectively.  The second uncertainty can be
removed by determining $\lambda_1$, as discussed in Sec.~II.  Other
applications, such as for nonleptonic decays, exclusive semileptonic decays and
$\bar B\to X_s\gamma$ photon spectrum were studied in Refs.~\cite{upsexp,LLMW}.

\begin{figure}[bt]
\centerline{\epsfysize=6.8cm\epsffile{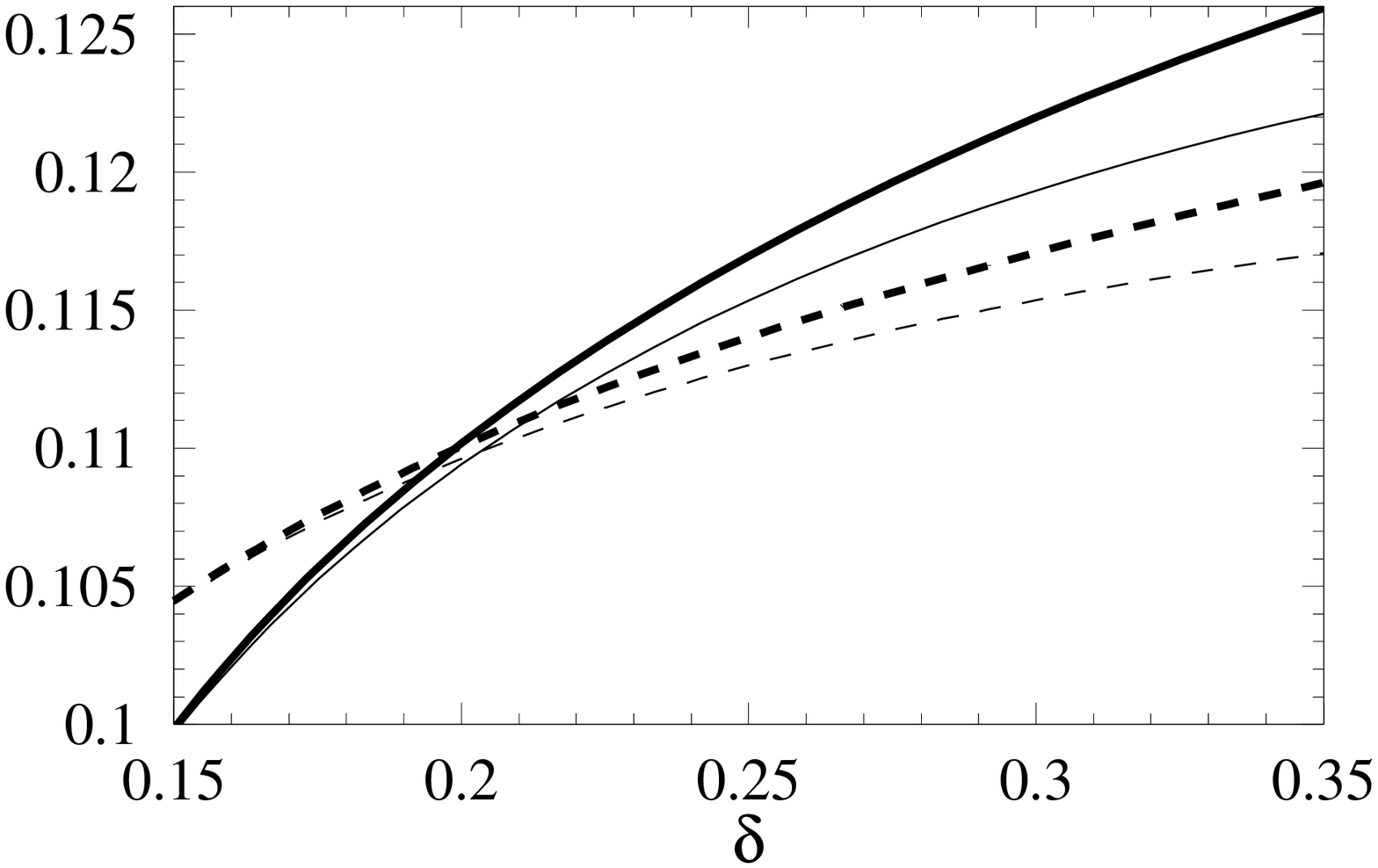}}
\caption[]{Prediction in the upsilon expansion at order $\epsilon$ (thick 
dashed curve) and $(\epsilon^2)_{\rm BLM}$ (thick solid curve) for 
$\overline{(1 - x_B)} |_{x_B > 1-\delta}$ defined in Eq.~(\ref{moment1}).
The thin curves show the $O_7$ contribution only.  (From Ref.~\cite{LLMW}.) }
\end{figure}

The most important uncertainty in this approach is the size of nonperturbative
contributions to $m_\Upsilon$ other than those which can be absorbed into the
$b$ quark mass.  By dimensional analysis the size of this correction is of
order $a^3\Lambda_{\rm QCD}^4$, where $a\sim1/(m_b \alpha_s)$ is the Bohr
radius of the $\Upsilon$.  Quantitative estimates, however, vary in a large
range, and it is preferable to constrain such effects from data.  The upsilon
expansion yields parameter free predictions for $\overline{(1 - x_B)} |_{x_B >
1-\delta}$ defined in Eq.~(\ref{moment1}).  The analog of Eq.~(\ref{beauty})
is~\cite{LLMW}
\begin{equation}\label{upsbeauty}
\overline{(1 - x_B)} \Big|_{x_B > 1-\delta} = 1 -
  {m_\Upsilon\over 2m_B} \left[ 1 + 0.011\epsilon + 0.019(\epsilon^2)_{\rm BLM}
  - \langle 1-x_b \rangle \Big|_{x_b > (2m_B/m_\Upsilon)(1-\delta)} \right] ,
\end{equation}
For $E_\gamma>2.1\,$GeV this relation gives 0.111, whereas the central value
from the CLEO data is around 0.093.  Fig.~4 shows the prediction for
$\overline{(1 - x_B)} |_{x_B > 1-\delta}$ as a function of $\delta$, both at
order $\epsilon$ and $(\epsilon^2)_{\rm BLM}$.  The perturbation expansion 
is very well behaved.  In Eq.~(\ref{upsbeauty}) nonperturbative contributions
to $m_\Upsilon$ other than those which can be absorbed into the $b$ quark mass
have been neglected.  If the nonperturbative contribution to $\Upsilon$ mass,
$\Delta_\Upsilon$, were known, it could be included by replacing $m_\Upsilon$
by $m_\Upsilon-\Delta_\Upsilon$. For example, $\Delta_\Upsilon = +100\,$MeV
increases $\overline{(1 - x_B)}$ by 7\%, so measuring $\overline{(1 - x_B)}$
with such accuracy will have important implications for the physics of
quarkonia as well as for $B$ physics.

\section{Conclusions}

To conclude, let me emphasize the main points, and indicate what data would be
important and useful in my opinion to address them:

\vspace{-9pt}
\begin{itemize} \itemsep=-4pt

\item Experimental determination of $\bar\Lambda$ and $\lambda_1$ from the
semileptonic $\bar B\to X_c e\bar\nu$ lepton energy and hadron mass spectra 
will reduce the theoretical uncertainties in $|V_{cb}|$ and $|V_{ub}|$. 
\hfil\break
(Need: Double tagged lepton spectrum with smaller errors.)

\item Photon energy spectrum in $\bar B\to X_s\gamma$ gives complimentary
information on $\bar\Lambda$ and $\lambda_1$, even in the presence of an 
experimental cut on the photon energy.
\hfil\break 
(Need: Spectrum with cut on $E_\gamma$ lowered; even a few hundred MeV can
reduce the uncertainties significantly.)

\item To distinguish $\bar B\to X_u e\bar\nu$ from $\bar B\to X_c e\bar\nu$,
cutting on the hadronic invariant mass is theoretically cleaner than cutting 
on the lepton energy. 
\hfil\break
(Need: Experimental cut on hadron mass as close to $m_D$ as possible, and a
precise determination of $\bar\Lambda$.)

\item The upsilon expansion is equivalent to using a short distance $b$ quark
mass, but it eliminates $m_b$ altogether from the theoretical predictions in
favor of $m_\Upsilon$ in a simple and consistent manner.  It raises several
interesting theoretical questions, and has many important applications.

\end{itemize}
\vspace{-9pt}

\acknowledgements 

I thank Adam Falk and Alan Weinstein for inviting me to give this talk, 
and Christian Bauer for providing Fig.~2.
%I~am grateful to British Airways for flying me out from London to LA
%directly instead of Chicago when O'Hare got shut down for days, thus 
%allowing me to deliver (and finish preparing\ldots) this talk.  
Fermilab is operated by Universities Research Association, Inc., under
DOE contract DE-AC02-76CH03000.


\begin{references}


\bibitem{sin2beta}
K. Pitts (CDF Collaboration), Fermilab Joint Experimental Theoretical Physics 
Seminar, February 5, 1999.

\bibitem{CGG}
J. Chay, H. Georgi and B. Grinstein, Phys. Lett. B247 (1990) 399;
M. Voloshin and M. Shifman, Sov. J. Nucl. Phys. 41 (1985) 120.

\bibitem{incl}
I.I. Bigi, N.G. Uraltsev and A.I. Vainshtein, Phys. Lett. B293 (1992) 430
[(E) Phys. Lett. B297 (1993) 477]; 
I.I. Bigi, M. Shifman, N.G. Uraltsev, and A. Vainshtein,
Phys. Rev. Lett. 71 (1993) 496.

\bibitem{MaWi}
A.V. Manohar and M.B. Wise, Phys. Rev. D49 (1994) 1310; 
B. Blok, L. Koyrakh, M.~Shifman and A.I. Vainshtein, 
Phys. Rev. D49 (1994) 3356; 
T. Mannel, Nucl. Phys. B413 (1994) 396.

\bibitem{SVlimit}
M.B. Voloshin and M.A. Shifman, Sov. J. Nucl. Phys. 47 (1988) 511.

\bibitem{SVdual}
C.G. Boyd, B. Grinstein, and A.V. Manohar, Phys. Rev. D54 (1996) 2081.

\bibitem{renormalon}
I.I. Bigi, M.A. Shifman, N.G. Uraltsev, A.I. Vainshtein, 
  Phys. Rev. D50 (1994) 2234;
M. Beneke and V.M. Braun, Nucl. Phys. B426 (1994) 301.

\bibitem{rencan}
M. Beneke, V.M. Braun, and V.I. Zakharov, Phys. Rev. Lett. 73 (1994) 3058;
M. Luke, A.V. Manohar, and M.J. Savage, Phys. Rev. D51 (1995) 4924;
M. Neubert and C.T. Sachrajda, Nucl. Phys. B438 (1995) 235.

\bibitem{gremmetal}
M. Gremm, A. Kapustin, Z. Ligeti, and M.B. Wise, 
  Phys. Rev. Lett. 77 (1996) 20;
M. Gremm and I. Stewart, Phys. Rev. D55 (1997) 1226.

\bibitem{Volo}
M.B. Voloshin, Phys. Rev. D51 (1995) 4934.

\bibitem{GK}
M. Gremm and A. Kapustin, Phys. Rev. D55 (1997) 6924.

\bibitem{FLSmass1}
A.F. Falk, M. Luke, and M.J. Savage, Phys. Rev. D53 (1996) 2491; 
Phys. Rev. D53 (1996) 6316.

\bibitem{FLSmass2}
A.F. Falk and M. Luke, Phys. Rev. D57 (1998) 424.

\bibitem{AZ}
A. Kapustin and Z. Ligeti, Phys. Lett. B355 (1995) 318.

\bibitem{LLMW}
Z. Ligeti, M. Luke, A.V. Manohar, and M.B. Wise, FERMILAB-PUB-99-025-T 
[hep-ph/9903305].

\bibitem{FLS}
A.F. Falk, M. Luke, and M.J. Savage, Phys. Rev. D49 (1994) 3367.

\bibitem{Bauer}
C. Bauer, Phys. Rev. D57 (1998) 5611.

\bibitem{KaNe}
A.L. Kagan and M. Neubert, Eur. Phys. J. C7 (1999) 5.

\bibitem{CLEObsg}
S. Glenn {\it et al.}, CLEO Collaboration, CLEO CONF 98-17.

\bibitem{SmVo}
B.H. Smith and M.B. Voloshin, Phys. Lett. B340 (1994) 176.

\bibitem{CLEOparams}
J. Bartelt {\it et al.}, CLEO Collaboration, CLEO CONF 98-21.

\bibitem{RoyPhD}
R. Wang, Ph.D. Thesis, University of Minnesota (1994).

\bibitem{Ryd}
See, e.g., The BaBar Physics Book (P.F. Harrison and H.R. Quinn, Eds.), 
Sec.~8.6.2, SLAC-R-504~(1998).

\bibitem{HQS}
N. Isgur and M.B. Wise, Phys. Lett. B232 (1989) 113; 
Phys. Lett. B237 (1990) 527.

\bibitem{IWprl}
N. Isgur and M.B. Wise, Phys. Rev. Lett. 66 (1991) 1130.

\bibitem{CLEObroad}
M. Zoeller, CLEO Collaboration, Talk presented at this Conference.

\bibitem{AlephCleo}
ALEPH Collaboration, D. Buskulic {\it et al.}, Z. Phys. C73 (1997) 601;
CLEO Collaboration, T.E. Browder {\it et al.}, Report no. CLEO CONF 96-2,
ICHEP96 PA05-077.

\bibitem{LLSW}
A.K. Leibovich, Z. Ligeti, I.W. Stewart, and M.B. Wise,
Phys. Rev. Lett. 78 (1997) 3995; Phys. Rev. D57 (1998) 308; 
Z.~Ligeti, hep-ph/9709396.

\bibitem{IWsr}
N. Isgur and M.B. Wise, Phys. Rev. D43 (1991) 319.

\bibitem{Luke}
M.E. Luke, Phys. Lett. B252 (1990) 447.

\bibitem{MNfact}
M. Neubert, Phys. Lett. B418 (1998) 173.

\bibitem{shapefn}
M. Neubert, Phys. Rev. D49 (1994) 4623; D49 (1994) 3392; 
A.F.~Falk, E.~Jenkins, A.V.~Manohar and M.B.~Wise, Phys. Rev. D49 (1994) 4553;
I.I. Bigi, M.A. Shifman, N.G. Uraltsev and A.I. Vainshtein, 
Int. J. Mod. Phys. A9 (1994) 2467;
R.D. Dikeman, M. Shifman, and N.G. Uraltsev, Int. J. Mod. Phys. A11 (1996) 571.

\bibitem{Misiak}
K. Chetyrkin, M. Misiak, and M. Munz, Phys. Lett. B400 (1997) 206;
M. Misiak and M. Munz, Phys. Lett. B344 (1995) 308.

\bibitem{match}
K. Adel and Y.P. Yao, Phys. Rev. D49 (1994) 4945; 
C. Greub and T. Hurth, Phys. Rev. D56 (1997) 2934.

\bibitem{fourquark}
A.J. Buras, M. Jamin, M.E. Lautenbacher, and P.H. Weisz, 
  Nucl. Phys. B370 (1992)~69.

\bibitem{matrixel}
C. Greub, T. Hurth, and D. Wyler, Phys. Rev. D54 (1996) 3350;
Phys. Lett. B380 (1996) 385.

\bibitem{1/mc2}
M.B. Voloshin, Phys. Lett. B397 (1997) 275; 
Z. Ligeti, L. Randall, and M.B. Wise, Phys. Lett. B402 (1997) 178; 
A.K. Grant, A.G. Morgan, S. Nussinov, R.D. Peccei, 
  Phys. Rev. D56 (1997) 3151;
G. Buchalla, G. Isidori, S.J. Rey, Nucl. Phys. B511 (1998) 594.

\bibitem{KLP}
A. Kapustin, Z. Ligeti, and H.D. Politzer, Phys. Lett. B357 (1995) 653.

\bibitem{FLW}
A.F. Falk, Z. Ligeti, M.B. Wise, Phys. Lett. B406 (1997) 225.

\bibitem{DiUr}
R.D. Dikeman and N.G. Uraltsev, Nucl. Phys. B509 (1998) 378;
I. Bigi, R.D. Dikeman, and N. Uraltsev, Eur. Phys. J. C4 (1998) 453.

\bibitem{oldmass}
V. Barger {\it et al.}, Phys. Lett. B251 (1990) 629;
J. Dai, Phys. Lett. B333 (1994) 212;

\bibitem{Vubmass}
R. Barate {\it et al.}, ALEPH Collaboration, CERN EP/98-067;
DELPHI Collaboration, contributed paper to the ICHEP98 Conference (Vancouver), 
paper 241; 
M. Acciarri {\it et al.}, L3 Collaboration, Phys. Lett. B436 (1998) 174.

\bibitem{ISGW}
N. Isgur {\it et al.}, Phys. Rev. D39 (1989) 799; 
N. Isgur and D. Scora, Phys. Rev. D52 (1995) 2783.

\bibitem{ACCMM}
G. Altarelli {\it et al.}, Nucl. Phys. B208 (1982) 365; 
A. Ali and I. Pietarinen, Nucl. Phys. B154 (1979) 519.

\bibitem{upsexp}
A.H. Hoang, Z. Ligeti, and A.V. Manohar, Phys. Rev. Lett. 82 (1999) 277;
Phys. Rev. D59 (1999) 074017 [hep-ph/9811239].

\bibitem{LSW}
M. Luke, M.J. Savage, an M.B. Wise, Phys. Lett. B343 (1995) 329; 
B345 (1995) 301; 
P. Ball, M. Beneke, and V.M. Braun, Phys. Rev. D52 (1995) 3929.

\bibitem{TvR}
T. Ritbergen, TTP-99-11 [hep-ph/9903226].

\bibitem{Upsmass}
A. Pineda and F.J. Yndurain, UB-ECM-PF-97-34 [hep-ph/9711287]; 
K. Melnikov and A. Yelkhovskii, TTP-98-17 [hep-ph/9805270].

\bibitem{ugo}
U. Aglietti and Z. Ligeti, Phys. Lett. B364 (1995) 75.

\bibitem{andre}
A.H. Hoang, M.C. Smith, T. Stelzer, and S. Willenbrock, 
  UCSD-PTH-98-13 [hep-ph/9804227].

\bibitem{Beneke}
M. Beneke, Phys. Lett. B434 (1998) 115.

\bibitem{SAZ}
A. Sinkovics, R. Akhoury, and V.I. Zakharov, UM-TH-98-08 [hep-ph/9804401].

\bibitem{BLM}
S.J. Brodsky, G.P. Lepage, and P.B. Mackenzie, Phys. Rev. D28 (1983) 228.


\end{references}
\end{document}